\newfam\scrfam
\batchmode\font\tenscr=rsfs10 \errorstopmode
\ifx\tenscr\nullfont
	\message{rsfs script font not available. Replacing with calligraphic.}
\else	\font\sevenscr=rsfs7 
	\font\fivescr=rsfs5 
	\skewchar\tenscr='177 \skewchar\sevenscr='177 \skewchar\fivescr='177
	\textfont\scrfam=\tenscr \scriptfont\scrfam=\sevenscr
	\scriptscriptfont\scrfam=\fivescr
	\def\scr{\fam\scrfam}
	\def\cal{\scr}
\fi
\newfam\msbfam
\batchmode\font\twelvemsb=msbm10 scaled\magstep1 \errorstopmode
\ifx\twelvemsb\nullfont\def\Bbb{\bf}
	\message{Blackboard bold not available. Replacing with boldface.}
\else	\catcode`\@=11
	\font\tenmsb=msbm10 \font\sevenmsb=msbm7 \font\fivemsb=msbm5
	\textfont\msbfam=\tenmsb
	\scriptfont\msbfam=\sevenmsb \scriptscriptfont\msbfam=\fivemsb
	\def\Bbb{\relax\expandafter\Bbb@}
	\def\Bbb@#1{{\Bbb@@{#1}}}
	\def\Bbb@@#1{\fam\msbfam\relax#1}
	\catcode`\@=\active
\fi
\font\eightrm=cmr8		\def\xrm{\eightrm}
\font\eightbf=cmbx8		\def\xbf{\eightbf}
\font\eightit=cmti8		\def\xit{\eightit}
\font\eighttt=cmtt8		\def\xtt{\eighttt}
\font\eightcp=cmcsc8
\font\eighti=cmmi8		\def\xold{\eighti}
\font\teni=cmmi10		\def\old{\teni}
\font\tencp=cmcsc10
\font\tentt=cmtt10
\font\twelverm=cmr12
\font\tencp=cmcsc10
\font\twelvecp=cmcsc10 scaled\magstep1
\font\fourteencp=cmcsc10 scaled\magstep2
\font\fiverm=cmr5		\def\trm{\fiverm}
\font\twelvemath=cmmi12
\font\fourteenmath=cmmi12 scaled\magstep1
\font\eightmath=cmmi8

\headline={\ifnum\pageno=1\hfill\else
{\eightcp Cederwall, von Gussich, Nilsson, Sundell, Westerberg: 
	``The Dirichlet Super-{\xit p}-Branes $\ldots$''}
		\dotfill{ }{\old\folio}\fi}
\def\makeheadline{\vbox to 0pt{\vss\noindent\the\headline\break
\hbox to\hsize{\hfill}}
	\vskip1.65\baselineskip}
\def\makefootline{\ifnum\foottest=1
	\baselineskip=.8cm\line{\the\footline}\global\foottest=0
	\fi
        }
\newcount\foottest
\global\foottest=0
\def\Footnote#1#2{${\,}^#1$\footline={\vtop{\baselineskip=10pt
        \hrule width.5\hsize\hfill\break
        \indent ${}^#1$ \vtop{\hsize=14cm\noindent\xrm #2}}}\foottest=1
        }
\newcount\refcount
\refcount=0
\newwrite\refwrite
\def\ref#1#2{\global\advance\refcount by 1
	\xdef#1{{\old\the\refcount}}
	\ifnum\the\refcount=1
	\immediate\openout\refwrite=\jobname.refs
	\fi
	\immediate\write\refwrite
		{\item{[{\xold\the\refcount}]} #2\hfill\par\vskip-2pt}}
\def\refout{\catcode`\@=11 
	\xrm\immediate\closeout\refwrite
	\vskip1.5\baselineskip
	{\noindent\twelvecp References}\hfill\vskip.5\baselineskip
	\parskip=.875\parskip 
	\baselineskip=.8\baselineskip
	\input\jobname.refs 
	\parskip=8\parskip \divide\parskip by 7
	\baselineskip=1.25\baselineskip 
	\catcode`\@=\active\rm}
\newcount\sectioncount
\sectioncount=0
\def\section#1#2{\global\eqcount=0
	\global\advance\sectioncount by 1
	\vskip\baselineskip\noindent
	\hbox{\twelvecp\the\sectioncount. #2\hfill}\vskip.3\baselineskip
	\xdef#1{\the\sectioncount}\noindent}
\newcount\appendixcount
\appendixcount=0
\def\appendix#1{\global\aeqcount=0
	\global\advance\appendixcount by 1
	\vskip\baselineskip\noindent
	\ifnum\the\appendixcount=1
	\hbox{\twelvecp Appendix A. #1\hfill}\vskip.3\baselineskip\fi
    \ifnum\the\appendixcount=2
	\hbox{\twelvecp Appendix B. #1\hfill}\vskip.3\baselineskip\fi
    \ifnum\the\appendixcount=3
	\hbox{\twelvecp Appendix C. #1\hfill}\vskip.3\baselineskip\fi\noindent}
\newcount\eqcount
\eqcount=0
\def\Eqn#1{\global\advance\eqcount by 1
	\xdef#1{{\old\the\sectioncount}.{\old\the\eqcount}}
		\eqno({\oldstyle\the\sectioncount}.
		{\oldstyle\the\eqcount})}
\def\eqn{\global\advance\eqcount by 1
	\eqno({\oldstyle\the\sectioncount}.{\oldstyle\the\eqcount})}
\def\multi{\global\advance\eqcount by 1}
\def\multieq#1#2{\xdef#1{{\old\the\eqcount#2}}
	\eqno{({\oldstyle\the\eqcount#2})}}
\newcount\aeqcount
\aeqcount=0
\def\AEqn#1{\global\advance\aeqcount by 1
	\ifnum\the\appendixcount=1 
		\xdef#1{\hbox{\xrm A}.{\old\the\aeqcount}}
		\eqno(\hbox{\xrm A}.{\oldstyle\the\aeqcount})\fi
	\ifnum\the\appendixcount=2
		\xdef#1{\hbox{\xrm B}.{\old\the\aeqcount}}
		\eqno(\hbox{\xrm B}.{\oldstyle\the\aeqcount})\fi
	\ifnum\the\appendixcount=3
		\xdef#1{\hbox{\xrm C}.{\old\the\aeqcount}}
		\eqno(\hbox{\xrm C}.{\oldstyle\the\aeqcount})\fi}
\def\aeqn{\global\advance\aeqcount by 1
	\ifnum\the\appendixcount=1 
		\eqno(\hbox{\xrm A}.{\oldstyle\the\aeqcount})\fi
	\ifnum\the\appendixcount=2
		\eqno(\hbox{\xrm B}.{\oldstyle\the\aeqcount})\fi
	\ifnum\the\appendixcount=3
		\eqno(\hbox{\xrm C}.{\oldstyle\the\aeqcount})\fi}
\parskip=3.5pt plus .3pt minus .3pt
\baselineskip=14pt plus .5pt minus .1pt
\lineskip=.5pt plus .05pt minus .05pt
\lineskiplimit=.5pt
\abovedisplayskip=19pt plus 4pt minus 2pt
\belowdisplayskip=\abovedisplayskip
\hsize=15cm
\vsize=20.5cm
\hoffset=1cm
\voffset.5cm
\def\/{\over}
\def\*{\partial}
\def\a{\alpha}
\def\b{\beta}
\def\c{\gamma}
\def\d{\delta}
\def\e{\varepsilon}

\def\g{\gamma}
\def\k{\kappa}
\def\l{\lambda}
\def\m{\mu}
\def\n{m}
\def\p{\phi}

\def\t{\theta}
\def\y{\eta}

\def\x{\xi}

\def\D{\Delta}
\def\O{\Omega}
\def\F{{\cal F}}
\def\tF{\tilde\F}

\def\G{\Gamma}
\def\L{{\cal L}}
\def\LL{\Lambda}

\def\R{{\Bbb R}}

\def\Z{{\Bbb Z}}

\def\N{{\Bbb N}}
\def\punkt{\,.}
\def\komma{\,,}
\def\.{.\hskip-1pt }

\def\-{\!-\!}
\def\+{\!+\!}
\def\={\!=\!}
\def\>{\!>\!}

\def\half{{\lower2pt\hbox{\eightrm 1}\/\raise2pt\hbox{\eightrm 2}}}
\def\frac#1{{\lower2pt\hbox{\eightrm1}\/\raise2pt\hbox{\eightrm#1}}}
\def\tfrac#1{{\lower2pt\hbox{\trm1}\/\raise2pt\hbox{\trm#1}}}
\def\genfrac#1#2{{\lower2pt\hbox{\eightrm#1}\/\raise2pt\hbox{\eightrm#2}}}

\def\tr{\hbox{\rm tr}}
\def\ie{{i.e.,}}
\def\eg{e.g\.}

\def\DBI{\hbox{Dirac--Born--Infeld}}
\def\dbi{{\hbox{\fiverm DBI}}}
\def\wz{{\hbox{\fiverm WZ}}}
\def\det{\hbox{\rm det}}
\def\II{\hbox{I\hskip-0.6pt I}}
\def\RR{R\hskip-0.6pt R}
\def\NSNS{N\hskip-0.6pt S\hskip.6pt-\hskip-.6pt N\hskip-0.6pt S}

\def\gel{\g_{11}}
\def\indab{{(\a\b}}
\def\indg{{\g)}}
\def\indgd{{\g)\d}}

\def\tG{\tilde\G}
\def\id{1\!\!1}

%
%
%
%

\null\vskip-1.5cm
\hbox to\hsize{\hfill G\"oteborg-ITP-96-14}
\hbox to\hsize{\hfill CTP-TAMU-59/96}
\hbox to\hsize{\hfill\tt hep-th/9611159}
\hbox to\hsize{\hfill November, 1996}

\vskip2.5cm
\centerline{\fourteencp The Dirichlet Super-{\fourteenmath p}-Branes} 
\vskip4pt
\centerline{\fourteencp in Ten-Dimensional Type \II A and \II B Supergravity}
\vskip\parskip
\centerline{\twelvecp}

\vskip1.2cm
\centerline{\twelverm Martin Cederwall\raise7pt\hbox{\xit a}, 
	Alexander von Gussich\raise7pt\hbox{\xit a}, 
	Bengt E.W. Nilsson\raise7pt\hbox{\xit a,b},} 
\vskip2pt
\centerline{\twelverm Per Sundell\raise7pt\hbox{\xit b}
	 and Anders Westerberg\raise7pt\hbox{\xit a}}

\vskip1.2cm
\centerline{\raise7pt\hbox{\xit a}\it Institute of Theoretical Physics}
\centerline{\it G\"oteborg University and Chalmers University of Technology }
\centerline{\it S-412 96 G\"oteborg, Sweden}

\vskip.4cm
\catcode`\@=11
\centerline{\tentt tfemc,tfeavg,tfebn,tfeawg@fy.chalmers.se}
\catcode`\@=\active

\vskip.8cm
\centerline{\raise7pt\hbox{\xit b}\it Center for Theoretical Physics}
\centerline{\it Texas A \& M University}
\centerline{\it College Station, Texas 77843, USA}

\vskip.4cm
\catcode`\@=11
\centerline{\tentt per@chaos.tamu.edu}
\catcode`\@=\active

\vskip2.5cm

\centerline{\bf Abstract}

{\narrower\noindent We give the full supersymmetric 
and $\k$-symmetric actions for the Dirichlet $p$-branes, including their
coupling to background superfields of ten-dimensional 
type \II A and \II B supergravity.\smallskip} 

\vfill

\eject

\def\nl{\hfill\break\indent}

\ref\CvGNW{M.~Cederwall, A.~von~Gussich, B.E.W.~Nilsson and A.~Westerberg,\nl
	{\xit ``The Dirichlet super-three-brane in type IIB supergravity''},
	{\xtt hep-th/9610148}.}
\ref\Aganagic{M.~Aganagic, C.~Popescu and J.H.~Schwarz,
	{\xit ``D-brane actions with local kappa symmetry''},
	{\xtt hep-th/9610249}.}
\ref\Polchinski{J. Polchinski, {\xit ``Dirichlet-branes and Ramond--Ramond 
	charges''},
	\nl Phys.~Rev.~Lett.~{\xbf 75} ({\xold1995}) {\xold4724} 
	({\xtt hep-th/9510017}).}
\ref\PolchinskiWitten{J.~Polchinski and E.~Witten,
	{\xit ``Evidence for heterotic--type I string duality''},
	\nl Nucl.~Phys.~{\xbf B460} ({\xold1996}) {\xold525} 
	({\xtt hep-th/9510169}).}
\ref\WittenIV{E.~Witten, {\xit ``Small instantons in string theory''},
	Nucl.~Phys.~{\xbf B460} ({\xold1996}) {\xold541} 
	({\xtt hep-th/9511030}).}
\ref\BST{E.~Bergshoeff, E.~Sezgin and P.K.~Townsend, 
	\nl {\xit ``Supermembranes and eleven-dimensional supergravity''},
	Phys.~Lett.~{\xbf B189} ({\xold1987}) {\xold75};
	\nl {\xit ``Properties of the eleven-dimensional supermembrane 
	theory''}, Ann.~Phys. {\xbf 185} ({\xold1988}) {\xold330}.}
\ref\Achucarro{A.~Ach\'ucarro, J.M.~Evans, P.K.~Townsend and D.L.~Wiltshire,
	{\xit ``Super p-branes''},
	\nl Phys.~Lett.~{\xbf B198} ({\xold1987}) {\xold441}.} 
\ref\BSTII{E.~Bergshoeff, E.~Sezgin and P.K.~Townsend, 
	\nl {\xit ``Super p-branes as gauge theories of volume-preserving
		diffeomorphisms''},
	Ann.~Phys.~{\xbf 199} ({\xold1990}) {\xold340}.}
\ref\DuffLu{M.J.~Duff and J.X.~Lu, 
	{\xit ``Type II p-branes: the brane-scan revisited''}, 
	\nl Nucl.~Phys.~{\xbf B390} ({\xold1993}) {\xold276} 
	({\xtt hep-th/9207060}).}
\ref\HoweSezgin{P.S.~Howe and E.~Sezgin, {\xit ``Superbranes''},
	{\xtt hep-th/9607227}.}
\ref\Witten{E.~Witten, {\xit ``String theory dynamics in various dimensions''},
	Nucl.~Phys.~{\xbf B443}
	({\xold1995}) {\xold85} ({\xtt hep-th/9503124}).}
\ref\Schwarz{J.H.~Schwarz, {\xit ``The power of M theory''},
	Phys.~Lett.~{\xbf B367} ({\xold1996}) {\xold97}
	({\xtt hep-th/9510086});\nl
	{\xit ``Lectures on superstring and M theory dualities''},
	{\xtt hep-th/9607201}.}	
\ref\Sen{A.~Sen, {\xit ``Unification of string dualities''},
	{\xtt hep-th/9609176}.}
\ref\Sezgin{E.~Sezgin, {\xit ``The M algebra''}, {\xtt hep-th/9609086}.}
\ref\Banks{T.~Banks, W.~Fischler, S.H.~Shenker and L.~Susskind,
	\nl{\xit ``M theory as a matrix model: a conjecture''},
	{\xtt hep-th/9610043}.}
\ref\BrinkHowe{L.~Brink and P.~Howe, {\xit ``Eleven-dimensional supergravity 
	on the mass-shell in superspace''},
	\nl Phys.~Lett.~{\xbf 91B} ({\xold1980}) {\xold384}.}
\ref\Cremmer{E.~Cremmer and S.~Ferrara, 
	{\xit ``Formulation of eleven-dimensional supergravity 
	in superspace''},\nl Phys.~Lett.~{\xbf 91B} ({\xold1980}) {\xold61}.}
\ref\Guven{R.~G\"uven, 
	{\xit ``Black p-brane solutions of D=11 supergravity theory''}, 
	Phys.~Lett.~{\xbf B276} ({\xold1992}) {\xold49}.}
\ref\DuffStelle{M.J.~Duff and K.S.~Stelle,
	{\xit ``Multimembrane solutions of D=11 supergravity''}
	Phys.~Lett.~{\xbf B253} (1991) 113.}
\ref\Hull{C.M.~Hull and P.K.~Townsend, 
	{\xit ``Unity of superstring dualities''}, 
	\nl Nucl.~Phys.~{\xbf B438} ({\xold1995}) {\xold109} 
	({\xtt hep-th/9410167}).}
\ref\HullII{C.M.~Hull, {\xit ``String dynamics at strong coupling''},
	Nucl.~Phys.~{\xbf B468} ({\xold1996}) {\xold113} 
	({\xtt hep-th/9512181}).}
\ref\TownsendII{P.K.~Townsend, {\xit ``p-brane democracy''},
	{\xtt hep-th/9507048}.}
\ref\Bergshoeff{E.~Bergshoeff, C.M.~Hull and T.~Ort\'{\i}n,
	{\xit ``Duality in the type-II superstring effective action''},
	\nl Nucl.~Phys.~{\xbf B452} ({\xold1995}) {\xold547}
	({\xtt hep-th/9504081}).}
\ref\Khuri{M.J.~Duff, R.R.~Khuri and J.X.~Lu, {\xit ``String solitons''},
	Phys.~Rep.~{\xbf 259} ({\xold1995}) {\xold213} 
	({\xtt hep-th/9412184}).}
\ref\Duff{M.J.~Duff, {\xit ``Strong/weak coupling duality from the dual
	string''},\nl Nucl.~Phys.~{\xbf B442}
	({\xold1995}) {\xold47} ({\xtt hep-th/9501030}).}
\ref\Behrndt{K.~Behrndt, E.~Bergshoeff and B.~Jansen,
	{\xit ``Type II duality in six dimensions''},
	\nl Nucl.~Phys.~{\xbf B467} ({\xold1996}) {\xold100} 
	({\xtt hep-th/9512152}).}
\ref\Aharony{O.~Aharony, {\xit ``String theory dualities from M theory''},
	Nucl.~Phys.~{\xbf B476} ({\xold1996}) {\xold470}
	({\xtt hep-th/9604103}).}
\ref\Berkooz{M.~Berkooz, R.G.~Leigh, J.~Polchinski, J.~H. Schwarz, N.~Seiberg
      	and E.~Witten,\nl 
      	{\xit ``Anomalies, dualities, 
		and topology of D=6 N=1 superstring vacua''},\nl
	Nucl.~Phys. {\xbf B475} ({\xold1996}) {\xold115} 
      	({\xtt hep-th/9605184}).}
\ref\Leigh{R.G. Leigh, {\xit ``Dirac--Born--Infeld action from Dirichlet sigma
	model''}, Mod.~Phys.~Lett.~{\xbf A4} ({\xold1989}) {\xold2767}.
		\vfill\eject}
\ref\Tseytlin{A.A.~Tseytlin, {\xit ``Self-duality of Born--Infeld action and 
	Dirichlet 3-brane of Type IIB superstring''},
	\nl Nucl.~Phys.~{\xbf B469} ({\xold1996}) {\xold51}
	({\xtt hep-th/9602064}).}
\ref\Gutperle{M.B.~Green and M.~Gutperle, {\xit ``Comments on 3-branes''},
	Phys.~Lett.~\xbf B377 \xrm ({\xold1996}) {\xold28}
	({\xtt hep-th/9602077}).}
\ref\Douglas{M.~Douglas, {\xit ``Branes within branes''},
	{\xtt hep-th/9512077}.}
\ref\GHT{M.B.~Green, C.M.~Hull and P.K.~Townsend, \nl {\xit ``D-brane 
	Wess--Zumino actions, T-duality and the cosmological constant''},
	{\xtt hep-th/9604119}.}
\ref\HoweWest{P.S.~Howe and P.C.~West, 
	{\xit ``The complete N=2, d=10 supergravity''},
	Nucl.~Phys.~{\xbf B238} ({\xold1984}) {\xold181}.} 
\ref\Carr{J.L.~Carr, S.J.~Gates Jr.~and R.N.~Oerter, 
	\nl{\xit ``D=10, N=2A supergravity in superspace''},
	Phys.~Lett.~{\xbf 189B} ({\xold1987}) {\xold68}.}
\ref\Townsend{P.K. Townsend, {\xit ``D-branes from M-branes''}, 
	Phys.~Lett.~{\xbf B373} ({\xold1996}) {\xold68} 
	({\xtt hep-th/9512062}).}
\ref\Schmidhuber{C. Schmidhuber, {\xit ``D-brane actions''}, 
	Nucl.~Phys.~{\xbf B467} ({\xold1996}) {\xold146}
	({\xtt hep-th/9601003}).}
\ref\CGMNW{M.~Cederwall, A.~von~Gussich, A.~Mikovi\'c, B.E.W.~Nilsson
	and A.~Westerberg, \nl{\xit ``On the \DBI\ action for D-branes''},
	Phys.~Lett.~B (to appear), {\xtt hep-th/9606173}, .}
\ref\Candiello{A.~Candiello and K.~Lechner, {\xit ``Duality in supergravity
	theories''},\nl Nucl.~Phys.~{\xbf B412} ({\xold1994}) {\xold479}
	({\xtt hep-th/9309143}).}
\ref\HoweSez{P.S.~Howe and E.~Sezgin,
	{\xit ``d=11,p=5''}, {\xtt hep-th/9611008}.}

\section\intro{Introduction}
\noindent In a previous paper [\CvGNW] 
the $\k$-symmetric Dirichlet three-brane action
including its coupling to a general type \II B on-shell supergravity 
background was 
constructed, and some key aspects for general D$p$-branes were anticipated.
Using essentially the same methods we here complete this construction by 
deriving all the $\k$-symmetric Dirichlet $p$-brane  
actions including their on-shell background couplings.
For even and odd $p$ this background is type \II A and type \II B 
supergravity, respectively.
The subject of supersymmetric D-branes has also been addressed in
ref.~[\Aganagic], where the form of the $\k$-transformations was
conjectured for general $p$ in a flat background. 

Given the important r\^ole played by D-branes in non-perturbative
string theory [\Polchinski,\PolchinskiWitten,\WittenIV], it is an urgent 
issue to obtain a fuller understanding of their inherent dynamics.
Progress in this area would also mean filling a gap in our present picture 
of the theory of extended objects [\BST,\Achucarro,\BSTII,\DuffLu,\HoweSezgin],
where D-branes generalize ordinary $p$-branes by containing
non-scalar world-volume fields. We also believe that such a detailed
understanding of the mechanisms at work will turn out to be valuable
when one addresses aspects of a more fundamental underlying theory,
M-theory [\Witten,\Schwarz,\Sen,\Sezgin,\Banks]. 
The low-energy limit of M-theory, eleven-dimensional
supergravity [\BrinkHowe,\Cremmer], contains extended objects, 
namely a membrane and a five-brane [\Guven,\DuffStelle], and we envisage 
that the analysis of these, and of the various string dualities 
[\Witten,\Schwarz,\Sen,\Hull-\Berkooz] 
(hopefully) explained by M-theory, will
benefit from the techniques presented in this paper. 
We will comment further on this in the last section. 

We start out by reviewing the bosonic D-brane actions in section {\old2}.
Section {\old3} discusses the supergravity backgrounds in which
the D-branes propagate, and introduces the algebraic constraints
necessary for the supergravity theories (thereby put on-shell)
and, as it turns out, also for the supersymmetry of the D-brane actions.
In the presentation of the result in section {\old 4} we treat the type 
\II A and type \II B branes in parallel, in order to stress the similarities
as far as the $\k$-symmetry is concerned. Here we verify that, given
the constraints of section {\old 3} on the fields in the \NSNS\ sector
(coupling to the kinetic term of the D-brane action)
the constraints in the RR sector (entering a Wess--Zumino term) may be
read off from $\k$-symmetry. Thus, consistent propagation of D-branes
demands a background solving the equations of motion of the appropriate
supergravity theory. In section {\old 5}, we solve the Bianchi identities
relevant to our discussion, and find that the solutions agree with the
constraints given in section {\old3}. This is the only instance where
essential differences between the \II A and \II B cases emerge.
In section {\old6}, we try to put the results of this paper in a somewhat
broader perspective, and comment on the possible applicability
of our techniques and of the mechanisms at work in the D-brane actions
for some unsolved related problems.
The paper ends by three appendices. In appendix A we describe our conventions
and notation. Appendix B treats the properties of the projection
matrix associated with $\k$-symmetry, and
finally, in appendix C we present the details of the 
proof of $\k$-symmetry that we left out in section {\old 4}.  

\section\Result{Preliminaries}A bosonic D$p$-brane is described 
by a target space and world-volume 
covariant action $I=I_{\dbi}+I_{\wz}$. 
The first term is the Dirac--Born--Infeld action [\Leigh]
$$
I_{\dbi}=\int_{M}d^{p+1}\x\,\L_{\dbi}\komma\Eqn\DBIaction
$$
where
$$\eqalign{
\L_{\dbi}&=-e^{-\p}\sqrt{-\det(e^{\tfrac{2}\p}g_{ij}
	+(F_{ij}-B_{ij}))}\phantom{\int}\cr
	&=-e^{{p-3\/4}\p}\sqrt{-\det(g_{ij}+e^{-\tfrac{2}\p}
	(F_{ij}-B_{ij}))}
	\punkt\phantom{\int}\cr}
\Eqn\DBIlagrangian
$$
Here $\x^{i}$ are the coordinates of the $(p+1)$-dimensional bosonic 
world-volume $M$, 
which is mapped by world-volume fields 
$X^{m}$ into ten-dimensional curved target space
with vielbein ${e_{m}}^{a}$. 
This embedding induces a world-volume 
metric $g_{ij}={e_{i}}^{a}{e_{j}}^{b}\y_{ab}$, where
${e_{i}}^{a}=\partial_{i}X^{m}{e_{m}}^{a}$
is the pull-back of the target space vielbein to the world-volume. 
The world-volume also carries an intrinsic abelian gauge field $A$ with
field strength $F=dA=\frac{2}d\x^{j}\wedge d\x^{i}F_{ij}$.\Footnote{\star}
{Here we have omitted the factor ${\a'\over2\pi}$ 
in front of {\xit F}. We have also left out the 
world-volume tension appearing in the DBI action.}  
A bosonic D$p$-brane thus has $(10-p-1)+(p-1)=8$ degrees of freedom.
Furthermore, $B_{ij}$ and $\p$ are the pullbacks of the target space
Neveu-Schwarz--Neveu-Schwarz (\NSNS) two-form and dilaton fields, 
respectively. The exponential of the dilaton defines the coupling of the 
theory. The dilaton dependence in (\DBIlagrangian) corresponds to an 
Einstein frame metric. Note that for $p=3$ the dilaton coupling affects 
only $\F\equiv F-B$.
This is related to the fact that the D3-brane is self-dual
under the SL(2;$\Z$) of type \II B [\Tseytlin,\Gutperle]. 
We will elaborate further on the r\^oles of the various fields appearing
in the action above in the following sections.

The second part of the action is the Wess--Zumino term [\Douglas,\GHT], 
containing the
couplings of the D$p$-brane to the target space Ramond--Ramond (\RR) fields. 
It can be written compactly as
$$
I_{\wz} = \int_M e^{\F}\!\!\wedge C\komma\Eqn\WZaction
$$
where the \RR\ $n$-form gauge potentials (pulled back to the world-volume) are 
collected in 
$$
C= \bigoplus_n C_{(n)}\punkt\Eqn\Csum
$$
The integration in (\WZaction) automatically selects the proper forms in the 
sum (\Csum) --- odd for type \hbox{\II A} and even for type \II B.

We now wish to construct the 
corresponding target superspace covariant and $\k$-symmetric 
D$p$-brane action (both local symmetries). The target superspace 
coordinates $Z^{M}$ are formed out of the $10$ bosonic $X^{m}$ and of
$32$ fermionic coordinates $\theta^\mu$ grouped
into two Majorana--Weyl spinors. In type \II A superspace these
two spinors carry opposite chiralities, while in type \II B
superspace they are of the same chirality which is conventionally chosen
to be positive. The implementation of manifest 
target superspace covariance for the action is rather straightforward; 
the prerequisites for this step will be presented in detail in the next 
section. The world-volume $M$, which is still bosonic,
is then mapped into the target superspace by 
fields $Z^{M}(\x)$. The resulting D$p$-brane theory will therefore
seemingly have $16$ fermionic on-shell degrees of freedom in the world-volume.
The r\^{o}le of the $\k$-symmetry to be treated in section {\old 4} 
is to reduce this number by half.

\section\SuperSpace{The {\twelvemath D}=10 supergravity background}The 
curved target superspace geometry is described by the vielbein
one-form $E^{A}=dZ^{M}{E_{M}}^{A}$ and the torsion two-form 
$$T^{A}=DE^{A}\equiv 
dE^{A}+E^{B}\!\wedge{\omega_{B}}^{A}\punkt\eqn
$$ 
The covariant derivative $D$ is defined in type \II A superspace using a 
Lorentz connection one-form ${\omega_{A}}^{B}$. Note that
$d$ has a right action; see appendix A. In the case of type \II B,
there is in addition a U(1) connection associated with the fact that the
scalars of type \II B supergravity live in the coset space 
SU(1,1)/U(1) (see \eg\ ref.~[\HoweWest]). 
The Lorentzian assumption amounts to the conditions
${\omega_{a}}^{\b}=0={\omega_{\a}}^{b}$. 
The Lorentzian field strength, i.e. the curvature two-form, is defined as 
$${R_{A}}^{B}=d{\omega_{A}}^{B}+{\omega_{A}}^{C}\wedge{\omega_{C}}^{B}\komma
\eqn
$$ 
which by the Lorentzian assumption obeys
$$
{R_{a}}^{\b}=0={R_{\a}}^{b}.
\Eqn\TheLorentzianAssumption
$$
We also have the ``first'' and ``second'' Bianchi identities:
$$\eqalign{
DT^{A}&=E^{B}\wedge{R_{B}}^{A}\komma\cr
D{R_{A}}^{B}&=0\punkt\cr}
\Eqn\TheFirstAndSecondBI
$$
This structure alone cannot yield a consistent on-shell
supersymmetric target space background; there is need for more
on-shell bosonic degrees of freedom to balance the excess of
fermions in the above field content after it has been constrained
in the fashion explained below. The additional on-shell bosons
required are supplied by the abelian 
super-field strengths --- the \NSNS\ super-three-form
$$
H_{(3)}=dB_{(2)}\eqn
$$ 
and the \RR\ super-$n$-forms\Footnote{\star}
{We will not treat the case of non-vanishing 
cosmological constant in type I\hskip-.4pt IA massive supergravity, \ie\ 
{\xit R}\lower2pt\hbox{\fiverm (0)}={\xit m}=\hskip-5pt/\hskip2pt0.
}
$$\eqalign{
R&=e^{B_{(2)}}\!\wedge d(e^{-B_{(2)}}\!\wedge C)\equiv\bigoplus_{n=1}^{10}
R_{(n)}\komma\cr
&\quad C\equiv\bigoplus_{n=0}^{9}C_{(n)}\punkt\cr}
\Eqn\TheFieldStrengths
$$ 
These field strengths obey the Bianchi identities
$$\eqalign{
dH_{(3)}&=0\komma\cr
e^{B_{(2)}}\!\wedge d(e^{-B_{(2)}}\!\wedge R)=dR-R\wedge H&=0\komma
\cr}
\Eqn\BianchiIdentities
$$
and they are invariant under the gauge transformations
$$\eqalign{
\delta B_{(2)}&=d\l\komma\cr
\delta C &= e^{B_{(2)}}\!\wedge d\m\punkt
\cr}
\Eqn\TargetSpaceGaugeTransf
$$
Note that the Bianchi identity (\BianchiIdentities) allows one to
nullify all even or odd forms in $R$.
The on-shell type \II A [\Carr] and \II B [\HoweWest] 
supergravities are then found
by truncating to even and odd $R$, respectively, and then imposing
the following 
constraints on the field strength components of canonical dimension 
less than or equal to $\half$:
$$\eqalign{
&\phantom{\hbox{\II A:\quad}}{T_{\a\b}}^{c}
	=2i\g_{\a\b}^{c}\komma \qquad {T_{a\b}}^{c}=0\komma\cr
&\hbox{\II A:\quad}{T_{\a\b}}^\g=\genfrac{3}{2}{\d_{(\a}}^\g\LL_{\b)}
	+2{(\gel )_{(\a}}^\g(\gel \LL)_{\b)}
	-\frac{2}{(\g_a)}_{\a\b}(\g^a\LL)^\g\cr
& \quad\qquad\qquad +(\g_a\gel )_{\a\b}(\g^a\gel \LL)^\g	
	+\frac{4}{(\g_{ab})_{(\a}}^\g(\g^{ab}\LL)_{\b)}\komma\cr
&\hbox{\II B:\quad}{T_{\a\b}}^\c=-{(J)_{(\a}}^\c(J\LL)_{\b)}
		+{(K)_{(\a}}^\c(K\LL)_{\b)}\cr
& \quad\qquad\qquad +\frac{2}(\g_aJ)_{\a\b}(\g^aJ\LL)^\c
		-\frac{2}(\g_aK)_{\a\b}(\g^aK\LL)^\c\komma\cr
&\cr
&\hbox{\phantom{\II A:\quad}} H_{\a\b\c}=0\komma\cr
&\hbox{\II A:\quad}H_{a\b\c}=-2i e^{\tfrac{2}\p}(\gel \g_{a})_{\b\c}\komma\cr
&\hbox{\phantom{\II A:\quad}}H_{ab\c}
	=e^{\tfrac{2}\p}(\g_{ab} \gel \LL)_{\c}\komma\cr
&\hbox{\II B:\quad}H_{a\b\c}=-2i e^{\tfrac{2}\p}( K\g_{a})_{\b\c}\komma\cr
&\hbox{\phantom{\II A:\quad}}H_{ab\c}
	=e^{\tfrac{2}\p}(\g_{ab}  K\LL)_{\c}\komma\cr
&\cr
&\hbox{\phantom{\II A:\quad}} R_{(n)\a\b\c A_{1}...A_{n-3}}=0\komma\cr
&\hbox{\II A:\quad}R_{(n)a_{1}...a_{n-2}\a\b}=2i\,e^{{n-5\/4}\p}
(\g_{a_{1}...a_{n-2}}(\gel )^{{n\/2}})_{\a\b}\komma\cr
&\hbox{\phantom{\II A:\quad}}R_{(n)a_{1}...a_{n-1}\a}
	=-\genfrac{{\xit n}-5}{2}e^{{n-5\/4}\p}
	(\g_{a_{1}...a_{n-1}}(-\gel )^{{n\/2}}\LL)_{\a}\komma\cr
&\hbox{\II B:\quad}R_{(n)a_{1}...a_{n-2}\a\b}=2i\,e^{{n-5\/4}\p}
	(\g_{a_{1}...a_{n-2}}K^{{{n-1}\/2}}I)_{\a\b}\komma\cr
&\hbox{\phantom{\II A:\quad}}R_{(n)a_{1}...a_{n-1}\a}
	=-\genfrac{{\xit n}-5}{2}e^{{n-5\/4}\p}
	(\g_{a_{1}...a_{n-1}}K^{{n-1\/2}}I\LL)_{\a}\punkt\cr
&\cr
&\hbox{\phantom{\II A:\quad}}\LL_{\a}=\frac{2}\*_{\a}\p
}
\Eqn\TheConstraints
$$
Here $K$ and $J$ are SO(2) matrices appearing in the real formulation
of type \II B supergravity; see appendix A for further discussion
and an explanation of our conventions for the $\g$-matrices.
We have included here the constraint relating the dimension-0
scalar superfield $\p$ containing the target space dilaton to the
superfield $\LL_\a$ whose leading component is the spinor of the
appropriate supergravity multiplet.

There is of course some arbitrariness in the choice of constraints.
The numerical coefficient in front of the dimension-0 component of the
torsion is free to choose, but then the normalization of the field strength
in the \DBI\ part of the action will fix the absolute value of the one
in front of the dimension-0 component of $H$, and also relate the
ones for the $R$'s to the normalization of the Wess--Zumino term. Once these
are chosen, and the relation between $\p$ and $\LL$ is fixed, no
freedom remains. Our conventions for $H$ differ by a sign from
those in ref.~[\CvGNW], which will also imply sign differences in
the projection matrix for the parameter of $\k$-symmetry.

The virtue of these constraints is that the 
Bianchi identities (\BianchiIdentities) and (\TheFirstAndSecondBI) and the 
Lorentzian assumption (\TheLorentzianAssumption) 
now turn from identities into
equations for the component fields of dimension 0 and higher, thus
reducing the enormous unconstrained field content precisely down to 
the field content of the on-shell type \II A and type \II B super-multiplets
respectively. Of course, one in particular 
has to check that these equations are
solved at dimensions $0$ and $\genfrac{1}{2}$ by (\TheConstraints).
This will be done in section 5. 
A more general analysis would reveal that 
the ``higher'' \RR\ field strengths $R_{(n)}$ with $n\geq5$ are
auxiliary in the sense that
their propagating degrees of freedom which sit in the
unconstrained  dimension-1 components $R_{(n)a_{1}...a_{n}}$
are related by Hodge duality to the propagating 
degrees of freedom
of the ``lower'' \RR\ field strengths $R_{(10-n)}$ contained
in analogous bosonic \hbox{$(10-n)$}-forms $r_{(10-n)}$ [\Douglas,\DuffLu].
In fact, in an on-shell background as we have here, 
one can consistently incorporate 
both a potential and its dual potential 
simultaneously, since the Bianchi identities and 
the field equations in such a situation are on equal footing.

To complete the picture of the target space background
we also include the
auxiliary dual \hbox{\NSNS} field strengths 
$$
\eqalign{
\hbox{\II A:\quad}&H_{(7)}=dB_{(6)}-\half C_{(1)}\wedge R_{(6)}
	+\half C_{(3)}\wedge R_{(4)}-\half C_{(5)}\wedge R_{(2)}\komma\cr
\hbox{\II B:\quad}&H_{(7)}=dB_{(6)}+\half C_{(0)}\wedge R_{(7)}
	-\half C_{(2)}\wedge R_{(5)}+\half C_{(4)}\wedge R_{(3)}
	-\half C_{(6)}\wedge R_{(1)}\komma\cr}
\eqn
$$ 
which are subject to the constraints
$$
\eqalign{
\hbox{\II A:\quad}&H_{a_{1}...a_{5}\a\b}
	=2ie^{-{\p\/2}}(\g_{a_{1}...a_{5}})_{\a\b}\komma\cr
&H_{a_{1}...a_{6}\a}=-e^{-{\p\/2}}(\g_{a_{1}...a_{6}}\LL)_\a\komma\cr
\hbox{\II B:\quad}&H_{a_{1}...a_{5}\a\b}
	=2ie^{-{\p\/2}}(\g_{a_{1}...a_{5}}K)_{\a\b}\komma\cr
&H_{a_{1}...a_{6}\a}=-e^{-{\p\/2}}(\g_{a_{1}...a_{6}}K\LL)_\a\komma\cr}\eqn
$$
and obey the Bianchi identities
$$
\eqalign{
\hbox{\II A:\quad}&dH_{(7)}+R_{(2)}\wedge R_{(6)}-\half R_{(4)}
\wedge R_{(4)}\komma\cr
\hbox{\II B:\quad}&dH_{(7)}+R_{(1)}\wedge R_{(7)}-R_{(3)}\wedge 
R_{(5)}=0\punkt\cr}\eqn
$$
Though not appearing explicitly in the Bianchi identities 
(\BianchiIdentities), this field strength plays a r\^{o}le
in the solution of the Bianchi identities for type \II B, as 
will be explained in more detail in section {\old 5}.

Whereas the auxiliary 
field content has previously been seen as an extravagance from
a target space point of view, its crucial function
in the construction of 
$\k$-symmetric $p$-brane actions is now well understood. 
There is only one exception,
namely the auxiliary \NSNS\ potential $B_{(6)}$ which does not enter the
construction of any of the $p$-brane actions that we know explicitly. 
The potential r\^{o}le
for $B_{(6)}$ in the construction of ($\k$-symmetric) 
world-volume actions will be discussed briefly in section~{\old 6}.

\section\TheResult{The Actions}As a consequence of the 
embedding of the bosonic world-volume $M$ into
target superspace, all target space structures are pulled back to 
$M$\Footnote{\star}{We follow the convention to use the same
notation for the super-forms in target space
as for their pull-backs to the world-volume (where they
become ordinary bosonic forms).}. 
The pull-back of the super-vielbein is given by
$$
E^{A}\equiv d\x^{i}{E_{i}}^{A}=d\x^{i}\partial_{i}Z^{M}{E_{M}}^{A}\punkt
\Eqn\PullBackVielBein
$$
Consequently, the pull-back of a super-$n$-form
$\O$ 
has the components
$$
\O_{i_1...i_n}={E_{i_{n}}}^{A_{n}}\cdots{E_{i_{1}}}^{A_{1}}
\O_{A_{1}...A_{n}}\punkt
\Eqn\PullBacknForm
$$
The pull-back of the Lorentz metric is given by
$$
g_{ij}={E_{i}}^{a}{E_{j}}^{b}\y_{ab}\punkt
\Eqn\PullBack 
$$
By insisting on manifest target space gauge invariance,
one is led to introduce the two-form field strength 
$$
\F=F-B_{(2)}\komma
\Eqn\TheGaugeInvField
$$
which is invariant under the target space transformation 
(\TargetSpaceGaugeTransf) combined with $\delta A=\l$.
We also note the gauge invariance of the Wess--Zumino term:
$$
\delta (e^{\F}\!\!\wedge C)=e^{\F}\!\!\wedge e^{B_{(2)}}\!\wedge
d\m=e^{dA}\!\wedge d\m=d(e^{dA}\m)\punkt
\Eqn\GaugeInvOfWZTerm
$$

Using this requisite we can now write the following
manifestly target superspace covariant and gauge invariant as well as 
world-volume covariant generalization of the action given
in section {\old 2}:
$$
I=I_{\dbi}+I_{\wz}=-\int_{M}d^{p+1}\x e^{{p-3\/4}\p}
\sqrt{-\det(g_{ij}+e^{-{\tfrac{2}\p}}\F_{ij})}+\int_{M} e^{\F}\!\!
\wedge C\punkt
\Eqn\TheTotalAction
$$
Formally, this action looks identical to the one in 
(\DBIaction)--(\WZaction). The crucial difference is of course that
the bosonic background fields have been replaced by their 
corresponding superfields.

Remarkably enough, the action (\TheTotalAction)
is now invariant under the following local variation 
$$\eqalign{
\delta_{\k}Z^{M}&=\k^{A}{E_{A}}^{M}\komma\quad\k^{a}=0\komma\cr
\delta_{\k}A&=i_{\k}B_{(2)}\komma
\cr}
\Eqn\TheKappaVariation
$$
provided that the spinorial variational parameter $\k$ obeys
$$
\G\k=\k\punkt
\Eqn\GammaKappaEqualsKappa
$$ 
Here the matrix $\G$ (acting in spinor space) is given by
$$
d^{p+1}\x\,\G=-{e^{\tfrac{4}(p-3)\p}\/\L_{\dbi}}
\exp\left(e^{-\tfrac{2}\p}\F\right)\wedge X Y|_{\rm vol}\komma
\Eqn\TheGammaMatrix
$$
with 
$$
X=\bigoplus_{n}\g_{(2n+q)}P^{n+q}
\Eqn\XDef
$$
and 
$$\eqalign{
\hbox{\II A:\quad} P&=\gel\komma\qquad Y=\id\komma\!\qquad q=1\komma\cr
\hbox{\II B:\quad} P&=K\komma\qquad\,\; Y=I\komma\qquad q=0\punkt
\phantom{\int}\cr}
\Eqn\PYq
$$
Here we have used the notation
$\g_{(n)}={1\/n!}d\x^{i_{n}}\wedge...\wedge d\x^{i_{1}}\g_{i_{1}...i_{n}}$,
where $\g_{i_{1}...i_{n}}$ is 
the pullback of the corresponding target space Dirac $\g$-matrix.
For further explanation of the notation, we again refer to appendix A.
This expression for $\G$ was conjectured for the \II B case in ref.~[\CvGNW].

The construction of the matrix $\G$ is one of the key steps in establishing
$\k$-symmetry of the D$p$-brane actions.
The crucial properties of $\G$ is that it squares to $\id$ and that 
the projection matrix $\frac{2}(\id+\G)$ has rank 16, i.e.~half the maximal 
value. The latter result follows immediately from the observation that
$\G$ is traceless and has eigenvalues $\pm1$.
 For a proof of the property $\G^2=\id$, we refer the 
reader to appendix B.

Hence, the $\k$-symmetry reduces the number of physical (real) fermionic 
degrees of freedom from 16 to 8, as required by supersymmetry. Of course,
this presupposes that fixing the world-volume diffeomorphism and gauge
invariances turns the target space spinors to spinors also on the
world-volume, as happens for the superstring and supermembrane. 
We have thus constructed the supersymmetric
Dirichlet $p$-branes for all possible values of $p$ in
type \II A and type \II B supergravity. 

Equipped with the matrix $\G$, we are now ready to establish the 
$\k$-invariance of the action (\TheTotalAction). 
The basic $\k$-variations of the world-volume fields given in 
(\TheKappaVariation) imply 
$$
\eqalign{\d_{\k} g_{ij}&=2{E_{(i}}^a{E_{j)}}^B \k^{\a}{T_{\a Ba}}
		\komma\cr
	\d_{\k}\F_{ij}&=-{E_j}^B{E_i}^A\k^{\a}H_{\a AB}\komma\cr
	\d_{\k}\p&=\k^{\a}\*_{\a}\p\punkt\cr}
\Eqn\gfpVariations
$$
The expression for $\d_\k\F_{ij}$ was obtained using the fact that
$$
\delta_{\k}\O=\{d,i_{\k}\}\O\komma
\Eqn\DeltaKappaForm
$$
when $\O$ is the pull-back of a 
super-form\Footnote{\star}{To be explicit,
$\delta_{\k}Z^{*}\O=Z^{*}\L_{\k}\O=Z^{*}\{d,i_{\k}\}\O$, where $Z$ is
the embedding of the world-volume in target superspace and  
$\L_{\k}$ is the target space Lie derivative
induced by the local target superspace vector field $\k$. We will not 
bother to write the $Z^{*}$ explicitly; see previous footnote.}.
Note that the definition of the 
$\k$-variation of the world-volume
one-form $A$ given in (\TheKappaVariation) is
chosen such as to cancel the term $d(i_\k B_{(2)})$
in the $\kappa$-variation of $B_{2}$.
The formula (\DeltaKappaForm) is of course 
also very useful when determining the variation of 
the WZ term in (\TheTotalAction), which is readily found to be
$$
\delta_{\k}I_{\wz}=\int_{M} e^{\F}\!\!\wedge i_{\k}R\punkt
\Eqn\DeltaKappaWZ
$$

At this point we can notice that since all 
field strengths, \NSNS\ as well as \RR , always appear
multiplied by $i_{\k}$, the conditions for 
$\k$-invariance will only constrain the components 
$R_{(n)A_{1}....A_{n-1}\a}$, i.e. the components of dimension less
than or equal to $\half$. 

It is less straightforward to obtain the $\k$-variation of the
DBI action. Using the matrix identity
$$
\delta\,\det M=\det M\,\tr(M^{-1}\delta M)\komma
\Eqn\MatrixVariation
$$ 
we find \hbox{\phantom{This text is here only for type setting reasons}}
$$
\d_{\G\k}I_{\dbi}=\int_M d^{p+1}\x\,\L_{\dbi}
	\left(\genfrac{{\eightmath p}-3}{4}
\d_{\G\k}\p+\half\tr\{(g+\tF)^{-1}(\d_{\G\k}g+\d_{\G\k}\tF)\}
\right)\komma
\Eqn\DBIvariation
$$
where we have defined $\tF\equiv e^{-{\p\/2}}\F$. 
Note here that we use $\G\k$ as the variational parameter. This is because 
when we insert the expression for $\G$ the factor $\L_{\dbi}$ in
(\DBIvariation) conveniently cancels.

Inserting in (\gfpVariations) the constraints for the torsion $T^A$, the 
\NSNS\ field strength $H_{(3)}$ and the dilaton $\p$ given in the
previous section, we obtain
$$\eqalign{
\d_{\G\k} g_{ij}&=4i(\bar E_{(i}\g_{j)}\,\G\k)\komma\cr
\d_{\G\k} \F_{ij}&=(-1)^q \left(4i(e^{\tfrac{2}\p}\,\bar E_{[i}\g_{j]} P\,\G\k)
		+e^{\tfrac{2}\p}(\bar\LL\g_{ij} P\,\G\k)\right)\komma\cr
\d_{\G\k}\p&=2\bar\LL\,\G\k\komma\cr}\Eqn\gfpWithConstr
$$
with $P$ and $q$ defined in (\PYq).
When inserting these expressions in (\DBIvariation), it is convenient to
consider the $\LL$-independent terms and the terms proportional to $\LL$
separately. Referring the reader to appendix C for the details of these 
somewhat lengthy but nevertheless illuminating calculations,
we only state the final results here: 
the contribution from the $\LL$-independent terms is
$$
\d^0_{\G\k}I_{\dbi}=\int_M e^\F\!\!\wedge\bigoplus_{n}
2i\,e^{\tfrac{4}(2n+q-4)\p}
(\bar\k\,\g_{(2n+q-1)}\,P^{n+q}Y\wedge E)\komma\Eqn\DeltaDBInoll
$$
while the $\LL$-dependent ones yield
$$
\d^{1/2}_{\G\k}I_{\dbi}= -\int_M e^\F\!\!\wedge \bigoplus_{n}
\,\half(2n+q-4)\,e^{\tfrac{4}(2n+q-4)\p}(\bar\LL\g_{(2n+q)}
\,P^{n+q}Y\k)\punkt
\Eqn\DeltaDBIhalv
$$
By comparing with (\DeltaKappaWZ) it is now straightforward to see that 
these expressions cancel the contributions to $\d_\k I_{\wz}$ coming from
the dimension-$0$ and dimension-$\half$ field strength components, 
respectively, precisely when these components satisfy the constraints 
given in (\TheConstraints).

We would like to stress that the result that the $\k$-variation of the
DBI action turns out to be a sum of differential forms appears to be
quite non-trivial.
It is even more gratifying to find that $\k$-invariance of the D$p$-brane 
action is not only possible, but also forces the background fields to
obey the appropriate supergravity equations of motion.

As mentioned in the introduction, the above result for the special case
$p$=3 was obtained earlier in ref.~[\CvGNW]. 
For $p$=2, the type \II A membrane action that we have derived here is 
equivalent to the previously known $\k$-symmetric type \II A membrane 
action, which by world-volume duality is equivalent to
the vertical dimensional reduction of the 
eleven-dimensional supermembrane [\Townsend,\Schmidhuber]. 
The latter action comes with an intrinsic world-volume metric $\g_{ij}$.
Solving the algebraic equations of motion for $\g_{ij}$ using the
formula for $p=2$ in ref.~[\CGMNW], we immediately recover
the supersymmetrized \DBI\ action (\TheTotalAction) (albeit
the rewriting of the $\G$-matrix is more tedious).
In fact, the proof of $\k$-symmetry of the
D3-brane action is not affected by the introduction of auxiliary
world-volume fields (the so called 1.5 order formalism). 
This observation generalizes immediately to the
general case presented here. 
\vfill\eject

\section\SolveBI{Bianchi Identities}Both chiral and non-chiral 
type \II\ supergravity in ten dimensions
have well-known formulations in superspace [\HoweWest,\Carr].
Therefore, it would in principle be sufficient to compare the 
dimension-$0$ and dimension-$\half$ components read off from $\k$-symmetry
and presented in eq.~(\TheConstraints) with the solutions of the
Bianchi identities given in these references. 

There are however several reasons to perform a
separate check. In type \II A, the constraints were originally written
in a formalism where Majorana spinor indices are divided
into dotted and undotted indices, while we have chosen the more compact
notation with a single Majorana index $\a$ (we refer to appendix A
for our spinor conventions). In type \II B, it is natural from a 
supergravity point of view to use a complex formalism, while the D-branes
couple in fundamentally different ways to \NSNS\ and \RR\ fields, so that
we are led to use a real formulation. It is even the case that the
fields that naturally enter the D-brane actions do not correspond to
real and imaginary parts of the same complex field.

One more reason why we actually have to solve the Bianchi identities is
that the higher D$p$-branes couple to higher antisymmetric tensor fields
not present in the ordinary formulations of the supergravities.
We thus need explicit control over the behavior of the tensor fields of
the dual type \II A and \II B supergravities.

We thus turn our attention to the list of constraints 
(\TheConstraints) and the 
relevant Bianchi identities, which when written out
in irreducible Lorentz components read as follows:
$$\eqalign{
&D_{[A}{T_{BC)}}^{D}+{T_{[AB}}^{E}{T_{|E|C)}}^{D}-{R_{[ABC)}}^{D}=0
	\komma\phantom{\int}\cr
&D_{[A}H_{BCD)}+\genfrac{3}{2}\,{T_{[AB}}^{E}H_{|E|CD)}=0
	\komma\phantom{\int}\cr
&D_{[A_{n+1}}R_{(n)A_{n}...A_{1})}+\frac{2}n\,{T_{[A_{n+1}A_{n}}}^{B}
R_{(n)|B|A_{n-1}...A_{1})}\phantom{\int}\cr
&\phantom{XXXX}\
	=\frac{6}n(n-1)R_{(n-2)[A_{n+1}...A_{4}}H_{A_{3}A_{2}A_{1})}\punkt\cr}
\Eqn\CompBI
$$
Here the brackets $[...)$ denote graded symmetrization of indices.
The constraints listed in (\TheConstraints) are not independent, 
since the Bianchi identities are a set of recursive relations
stepping upwards in dimension so that 
components of dimension d are expressed in terms of 
components of dimension d-$\half$ and below. 
The canonical dimensions of the torsion components and the components
of a field strength super-$n$-form $\O$ are given by:
$$\eqalign{
\hbox{d}({T_{AB}}^{C})&=\hbox{d}(A)+\hbox{d}(B)-\hbox{d}(C)\komma\cr
\hbox{d}(\O_{A_{1}...A_{n}})&=\hbox{d}(A_{1})+\cdots 
	+\hbox{d}(A_{n})+1-n\komma\cr}
\Eqn\CanonicalDimensions
$$
where $\hbox{d}(\a)=\hbox{d}(D_{\a})=\half$ 
and $\hbox{d}(a)=\hbox{d}(D_{a})=1$. 
All tensor fields of negative canonical dimension 
of course have to vanish,
while the dimension-$0$ components are fixed by Lorentz invariance up
to arbitrary super-scalar coefficients. These coefficients are related
through the
dimension-$0$ Bianchi identities in (\CompBI) to various powers of one 
overall (dimensionless) dilaton superfield coupling $e^{-\tfrac{2}\p}$.
For the torsion there is an additional degree of freedom
corresponding to the higher components in the
superfield expansion of the local super-Lorentz parameters 
that allows one to impose a conventional constraint on its
dimension-$\frac{2}$ and $1$ components. We have chosen the
conventional constraints
to be ${T_{a\b}}^{c}=0={T_{ab}}^{c}$. 
Once the
dimension-$0$ constraints and the conventional constraint
have been imposed, there is nothing else to be done other than to
compute the consequences for the higher-dimensional components.
The resulting structure is extremely rigid, in that 
the same components typically get determined over and over again.
If one would march on to higher canonical dimensions one would
at dimension $1$ find the Hodge duality relation mentioned in
section {\old 3}.
At dimension $\genfrac{3}{2}$ and $2$ one would find the spin-$\genfrac{3}{2}$
and spin-$1$ equations of motion, respectively. The spin-$2$
equations of motion sit at dimension $3$, while the dimension-$\genfrac{5}{2}$
equations yield closure of the supersymmetry algebra, i.e., express
the variation of the spin-$2$ field in terms of the spin-$\genfrac{3}{2}$
field.
For our purposes, however, it is sufficient to only solve the
Bianchi identities up to the dimension-$\frac{2}$ level, i.e. to 
check that (\TheConstraints) solves the Bianchi identities 
(\CompBI) at dimensions $0$ and $\half$.
In the type \II A case this procedure is quite straightforward
using only the cyclic Fierz identity in appendix A.

For the tensor fields of type \II B, there is an additional
complication due to the fact that the scalar fields sit in the coset
SL(2;$\R$)/U(1). There is a U(1) gauge field present, and different
fields carry different U(1) charge. In our real formalism, this will
be seen as an SO(2) connection mixing the \NSNS\ and \RR\
three-forms and seven-forms. 
Also, any object carrying a spinor index will carry
an SO(2) charge. The SO(2) action on spinor indices can be understood
as the adjoint action of the matrix $I$, so that $I$ itself and the
identity matrix are invariant, while $J$ and $K$ are rotated into
each other. Since the U(1) connection is a part of an SL(2;$\R$) connection,
it is not consistent to set it to zero, even though it is pure gauge;
rather one has to solve the Maurer--Cartan equation for SL(2;$\R$). This
is done for our real formalism in ref.~[\CvGNW], and once the gauge
choice is performed, one can integrate the Maurer--Cartan equations
to obtain two real fields, the dilaton and the \RR\ zero-form potential.
The gauge choice we perform makes the U(1) connection proportional
to the \RR\ one-form field strength, as explained in ref.~[\CvGNW].

The three-form and seven-form field strengths that enter the D$p$-brane
actions for odd $p$ do not carry any U(1) charge, so the covariant
derivatives in their Bianchi identities can be replaced by ordinary
exterior derivatives. Nevertheless, since the vielbeins stripped off
to obtain the dimension-$0$ and dimensions-$\half$ components do carry
charge through their spinor indices, it is important to note that the
Bianchi identities for these components require inclusion of connection
terms. The gauge choice performed is such that the connection enters
the Bianchi identities for $R_{(3)}$ and $R_{(7)}$ but not for $H$.
All other tensor fields are neutral under U(1).
Once this detail is taken into account the type \II B
case is quite straightforward.
The main Fierz identity needed is the cyclic identity of 
Appendix A.

\section\Conclusions{Conclusions and comments}It is worth 
emphasizing again that it is not at all a priori obvious 
that the $\k$-variation of the \DBI\ action can be written as a wedge products
of forms, allowing it to be cancelled by the variation of a
Wess--Zumino term. We have not investigated whether the class
of Lagrangians with this property contains more members than the present
example, but we are convinced that a check order by order
in the invariants of the matrix $\F_{ij}$ would not leave much room
for any deviations from the \DBI\ expression.

We would like to recapitulate what is known about supersymmetric extended
objects and their places in supergravity, string theory and M-theory.
When the D-brane actions are understood,
the only missing cases in string theory are the solitonic 
type \II A and \II B five-branes, and the heterotic five-brane, 
which is T-dual to the
well-understood type I five-brane. 
Both of the type \II\ five-branes couple to the \NSNS\ six-form potential in 
the Wess--Zumino term, but otherwise they seem to be very dissimilar. 
The type \II B
five-brane is simply obtained from the D5-brane by performing a duality
transformation on the background fields (which in particular reverts the
sign of the dilaton), and thus couples to a modified \RR\ two-form potential
in the kinetic (actually \DBI) term. Its construction is thus in practice
performed already in this paper.
The type \II A five-brane, on the 
other hand, will descend from the eleven-dimensional five-brane by
vertical dimensional reduction, and couples to the \RR\ three-form potential
in the kinetic term. Essentially, the missing information about
supersymmetric extended objects (in Minkowski signature) lies in 
the eleven-dimensional five-brane, which contains a world-volume
two-form potential with self-dual field strength, coupling in the kinetic term
to the three-form potential of eleven-dimensional supergravity,
and a Wess--Zumino coupling to the six-form potential of the
dual supergravity [\Candiello]. However, for recent progress in finding 
a formulation using 
the embedding formalism, see ref.~[\HoweSez]. We are quite optimistic about
the possibilities of obtaining further information 
 by applying the techniques
developed in the present paper. 

Type \II A supergravity is of course closely related to eleven-dimensional
supergravity, and it is plausible that all the structures encountered
will receive an explanation in terms of M-theory. While 
the contents of this statement are clear for $p$=1, 2, 4 and 5, 
it is not at all obvious what the r\^oles of
the D0- and D6-branes are. In ten dimensions they are dual to each other
(in the sense that they couple to dual field strengths), but from
an eleven-dimensional perspective the supergravity theory
(or its dual formulation) contains no fields of appropriate
rank they could couple to. The possible formulation of M-theory as
a composite zero-brane model [\Banks] is suggestive, and may shed light on
this puzzle.

\vskip3mm
{\tencp Acknowledgments}
\vskip-\parskip

B.E.W.N is grateful to the Center for Theoretical Physics at 
Texas A\&M University for its kind hospitality.
P.S. would like to thank M.J.~Duff, H.~Lu, C.N.~Pope and E.~Sezgin for 
enlightening comments 
related to this work. 
\vfill\eject

\appendix{Superspace Conventions and Fierz Identities}We work 
with a mostly positive Lorentz metric $\eta_{ab}=\hbox{diag}\,(-1,1,...,1)$ 
and Dirac
$\gamma$-matrices obeying $\{\g_{a},\g_{b}\}=2\eta_{ab}$. 
The unit-normalized $\gamma$-matrices $\g_{a_{1}...a_{n}}$ are defined by
$$
\g_{a_{1}...a_{n}}=\g_{[a_{1}}\cdots\g_{a_{n}]}\komma
\AEqn\DefOfASGammaMatrix
$$
where $[a_{1}...a_{n}]$ implies antisymmetrization with weight 1.
We frequently collect these objects in forms $\g_{(n)}$.
The chirality matrix $\g_{11}\equiv \g_{0}\g_{1}\cdots\g_{9}$ squares
to $+1$. 
The indices of a spinor $\psi^{\a}$ and a bispinor ${M^{\a}}_{\b}$ 
are (e.g. a $\gamma$-matrix) lowered and raised by means of ``natural''
matrix multiplication with the antisymmetric 
charge conjugation matrix $C_{\a\b}=-C_{\b\a}$ and its
inverse $C^{\a\b}\equiv (C^{-1})^{\a\b}$:
$$\eqalign{
C_{\a\b}C^{\b\g}&={\d_{\a}}^{\g}\komma\cr
\psi^{\a}&=C^{\a\b}\psi_{\b}\komma\cr
\psi_{\a}&=C_{\a\b}\psi^{\b}\komma\cr
{M_{\a}}^{\b}&=C_{\a\g}{M^{\g}}_{\d}C^{\d\b}\punkt\cr}
\AEqn\RaiseNLower
$$
We work with real Majorana spinors $\psi$ obeying
$$
\bar{\psi}=\psi^TC\punkt
\AEqn\ChargeConj
$$
These conventions minimize the number of minus-signs
from ``see-sawing'':
$$
\eqalign{
\bar{\psi}M\chi&\equiv \bar{\psi}_{\a}{M^{\a}}_{\b}\chi^{\b}=
\psi^{\a}M_{\a\b}\chi^{\b}\komma\cr
M_{\a\b}{M^{\prime\;\b}}_{\g}&={M_{\a}}^{\b}{M'}_{\b\g}\punkt\cr}
\AEqn\NoSeeSawSign
$$
As is well known, the $\gamma$-matrix $(\g_{a_{1}...a_{n}})_{\a\b}$
is antisymmetric when $n=0,3,4,7,8$ and symmetric when $n=1,2,5,6,9,10$.

In type \II A superspace the two $16$-component Majorana--Weyl
spinorial coordinates of opposite chirality form a
$32$-component Majorana spinor $\t^{\a}$, $\a=1,..,32$, acted
on (in the Weyl representation) by real block-off-diagonal 
Dirac $\g_{a}$-matrices, whose off-diagonal
$16$ by $16$ blocks are real Majorana $\sigma_{a}$-matrices, and a diagonal 
$\g_{11}$-matrix. 
A complete set of matrices for the product of two Majorana spinors is
$\gel$, $\g_{(1)}$, $\g_{(1)}\gel$, $\g_{(2)}$, $\g_{(4)}\gel$  
and $\g_{(5)}$ (symmetric) and $C$, $\g_{(2)}\gel$, $\g_{(3)}$, 
$\g_{(3)}\gel$ and $\g_{(4)}$ (antisymmetric).
The higher $\g$'s obey $\g_{(10-n)}=*\g_{(n)}\gel$. 

In type \II B superspace the two 16-component Majorana--Weyl
spinorial coordinates of equal chirality form a 32-component
spinor $\t^{\a}$, $\a=1,..,32$, whose 
$\a$-index is a composite index representing 
the tensor product of a Majorana--Weyl index
and an SO(2) index. This index is acted on by 
the real Majorana $\sigma_{a}$-matrices, and the real SO(2) matrices 
$\id$ and $I,J,K$ defined by $I^{2}=-\id$, $J^{2}=K^{2}=\id$, 
$IJ=K$ and $I,J,K$ anticommuting. The latter behave as 
generators for SL(2;$\R$), or, equivalently, as the imaginary split
quaternionic units (a convenient basis, though not necessary for
any calculations, is found in ref.~[\CvGNW]). 
A complete set of matrices for the tensor product $32\times32$ consists of
$\g_{(1)}$, $\g_{(1)}J$, $\g_{(1)}K$, $\g_{(3)}I$, $\g_{(5)}$, $\g_{(5)}J$
and $\g_{(5)}K$ (symmetric) together with $\g_{(1)}I$, $\g_{(3)}$, 
$\g_{(3)}J$, $\g_{(3)}K$ and $\g_{(5)}I$ (antisymmetric). We have
$\g_{(10-n)}=*\g_{(n)}$.

The basic cyclic Fierz identities are given by:
$$
\eqalign{
\hbox{\II A:}\quad&(\g^a)_\indab(\g_a)_\indgd=
	-(\g^a\gel)_\indab(\g_a\gel)_\indgd\komma\cr
\hbox{\II B:}\quad&(\g^a)_\indab(\g_a)_\indgd=-(\g^aJ)_\indab(\g_aJ)_\indgd
	=-(\g^aK)_\indab(\g_aK)_\indgd\punkt\cr}
\AEqn\CyclicFierzes
$$
Repeated use of these identities suffice for a quite direct 
verification of almost
all the Bianchi identities of section \SolveBI. Only for the higher
type \II B field strengths there seems to be a need for more general 
Fierz identities. The identity used in the dimension-$\half$
Bianchi identity for $R_{(7)}$ is
$$
\eqalign{
&(\g^a)_\indab{(\g_a\g_{a_{1}\ldots a_{5}})_\indg}^{\d}\cr
&\phantom{XXXX}-\genfrac{5}{2}(\g_{a_1})_\indab{(\g_{a_2\ldots a_5})_\indg}^\d
	+\genfrac{5}{2}(\g_{a_1}J)_\indab{(\g_{a_2\ldots a_5}J)_\indg}^\d
       	+\genfrac{5}{2}(\g_{a_1}K)_\indab{(\g_{a_2\ldots a_5}K)_\indg}^\d\cr
&\phantom{XXXX}-5(\g_{a_1a_2a_3}I)_\indab{(\g_{a_4 a_5}I)_\indg}^\d\cr
&\phantom{XXXX}-\half(\g_{a_1\ldots a_5})_\indab{\d_\indg}^\d
	+\half(\g_{a_1\ldots a_5}J)_\indab{J_\indg}^\d
      	+\half(\g_{a_1\ldots a_5}K)_\indab{K_\indg}^\d=0\cr}
\AEqn\FierzExample
$$
(antisymmetrization in $a_1\ldots a_5$ is understood),
and it can be verified by explicitly tracing with all the symmetric 
matrices. The Bianchi identity of $R_{(9)}$ involves the Fierz identity
$$
\eqalign{
&(\g^a)_\indab{(\g_a\g_{a_1\ldots a_7})_\indg}^\d\cr
&\phantom{XX}-7(\g_{a_1})_\indab{(\g_{a_2\ldots a_7})_\indg}^\d
	+\genfrac{7}{4}(\g_{a_1}J)_\indab{(\g_{a_2\ldots a_7}J)_\indg}^\d
	+\genfrac{7}{4}(\g_{a_1}K)_\indab{(\g_{a_2\ldots a_7}K)_\indg}^\d\cr
&\phantom{XX}+\genfrac{21}{4}
		(\g_{a_1\ldots a_5}J)_\indab{(\g_{a_6a_7}J)_\indg}^\d
	+\genfrac{21}{4}(\g_{a_1\ldots a_5}K)_\indab{(\g_{a_6a_7}K)_\indg}^\d
	-(\g_{a_1\ldots a_7}I)_\indab{I_\indg}^\d=0\punkt\cr
}\aeqn
$$

We use the superspace conventions in which a super-$n$-form $\O_{(n)}$ is
expanded as 
$$
\O_{(n)}={1\/n!}E^{A_{n}}\wedge\cdots\wedge E^{A_{1}}\O_{A_{1}...A_{n}}
\punkt\AEqn\SupernForm
$$
The exterior derivative 
$$
d=E^{A}\*_{A}\equiv dZ^{M}\partial_{M}
\AEqn\ExtDerivative
$$ 
has a right action on superforms:
$$\eqalign{
d\O_{(n)}&={1\/n!}dZ^{M_n}\wedge...\wedge dZ^{M_1}\wedge dZ^N \*_N
\O_{M_1...M_n}\phantom{\int}\cr
&={1\/n!}E^{A_{n}}\wedge\cdots\wedge E^{A_{1}}\wedge E^{A}\left(
D_{A}\O_{A_{1}...A_{n}}+\frac{2}n\, {T_{A A_1}}^{B}\O_{B A_2...A_n}\right)
\komma\phantom{\int}\cr 
d(\O_{(m)}\wedge\O_{(n)})&=\O_{(m)}\wedge d\O_{(n)}+
(-1)^{m}d\O_{(m)}\wedge\O_{(n)}\punkt\phantom{\int}\cr}
\AEqn\dActsFromRight
$$
The right action of the 
interior product $i_{V}$, where $V$ is a supervector field, is
$$\eqalign{
i_{V}\O_{(n)}&={1\/(n-1)!}E^{A_{n}}\wedge\cdots\wedge E^{A_{2}}
V^{A}\O_{AA_{2}...A_{n}}\komma\cr
i_{V}(\O_{(m)}\wedge\O_{(n)})&=\O_{(m)}\wedge i_{V}\O_{(n)}+
(-1)^{m}i_{V}\O_{(m)}\wedge\O_{(n)}\punkt\cr}
\AEqn\InnerDerivative
$$

\appendix{Proof of $\G^{2}=\id$}In this appendix 
we will prove that the matrix $\G$ defined in 
(\TheGammaMatrix) squares to $\id$. 
In order to
simplify the notation somewhat we will only consider the type \II B case ---
the IIA case follows analogously. 

For the type \II B D$p$-branes we can write (\TheGammaMatrix) as
$$
d^{p+1}\x\,\G={1\/\sqrt{-\det(g+\tF)}}\,e^{\tF}\!\wedge 
X I|_{\rm vol}\komma
\AEqn\IIBTheGammaMatrix
$$
with
$$
X=\bigoplus_{n\in\N}\g_{(2n)}K^{n}\punkt
\aeqn
$$
Hence, if we define ${\tG}=\sqrt{-\det(g+\tF)}\G$, we have to show that 
${\tG}^2=-\det(g+\tF)$. In dimension $p+1=2\n  $ the square of $\tG$ is
$$
\eqalign{\tG^2=&-\e^{i_{1}...i_{2\n }}\e^{j_{1}...j_{2\n }}\!\!\!\!\!\!\!\!\!\!
\sum_{k,r=0;\, k+r\,\hbox{\xrm{even}}}^{\n }\!\!\!\!\!\!\!\!(-1)^{r}{1\/(2k)!\,
(2r)!}\g_{i_{1}...i_{2k}}\g_{j_{1}...j_{2r}}\cr
&\qquad\times{1\/(\n -k)!\,(\n -r)!\,2^{2\n -k-r}}
\tF_{i_{2k+1}i_{2k+2}}...\tF_{i_{2\n -1}i_{2\n }}
\tF_{j_{2r+1}j_{2r+2}}...\tF_{j_{2\n -1}j_{2\n }}\komma\cr}
\AEqn\GammaSquare
$$
where the overall sign is due to $I^2=-\id$, and the factor $(-1)^{r}$ comes
from the fact that 
$IK^{r}=(-1)^{r}K^{r}I$. Furthermore, due to the symmetry in the 
$i$- and $j$-indices the sum $k+r$ will necessarily be even, implying that we
have only even powers in $\tF$. The $\g$-matrix
products can be expanded as
$$
\g_{i_{1}...i_{2k}}\g_{j_{1}...j_{2r}}=\g_{i_{1}...i_{2k}j_{1}...j_{2r}}+...+
c_q^{k,r}g_{[j_{1}...j_{2q}}^{[i_{1}...i_{2q}}
{\g^{i_{2q+1}...i_{2k}]}}_{j_{2q+1}...j_{2r}]}+...\komma
\AEqn\GammaProduct
$$
where 
$$
\eqalign{c_q^{k,r}&=(-1)^{q}{(2k)!\,(2r)!\/(2k-2q)!\,(2r-2q)!\,(2q)!}
\komma\cr
g_{j_{1}...j_{2q}}^{i_{1}...i_{2q}}&=g_{[i_{1}|j_{1}|}...g_{i_{2q}]j_{2q}}
\komma\cr
{\g^{i_{2q+1}...i_{2k}}}_{j_{2q+1}...j_{2r}}&=
\g_{i_{2q+1}...i_{2k}j_{2q+1}...j_{2r}}\punkt\phantom{\int}\cr}
\AEqn\GammaProductCoefficients
$$
For the same reason as above only even powers of the metric appear 
in the expansion (\GammaProduct). Note that the super-script indices
on $g$ and $\g$ have been introduced here only for notational reasons and
should not be interpreted as having been raised by $g^{-1}$.

In order to get the factor $-\det(g+\tF)$ we sum over all terms in
$\tG^2$ proportional to the identity matrix: 
$$
\eqalign{\tG^2|_{\id}=-\e^{i_{1}...i_{2\n }}e^{j_{1}...j_{2\n }}
\sum_{k=0}^{\n }{1\/(2k)!}&g^{i_{1}...i_{2k}}_{j_{1}...j_{2k}}
{1\/(\n -k)!\,(\n -k)!\,2^{2\n -2k}}\cr
&\quad\times\tF_{i_{2k+1}i_{2k+2}}...\tF_{i_{2\n -1}i_{2\n }}
\tF_{j_{2k+1}j_{2k+2}}...\tF_{j_{2\n -1}j_{2\n }}\punkt\cr}
\AEqn\GammaProductSurviving
$$
We move the indices between $\tF$'s using the
identity $S_{[i_{1}...i_{2\n +1}]}=0$. The result is schematically
$$
\e^{i}\e^{j} (g^{k})_{ij}
(\tF^{\n -k})_{ii}
(\tF^{\n -k})_{jj}=
{2^{\n -k}(\n -k)!\/(2\n -2k)-1)!!}\e^{i}\e^{j} 
(g^{k})_{ij}(\tF^{2\n -2k})_{ij}\komma
\AEqn\SchoutenGammaZero
$$
which inserted in (\GammaProductSurviving) gives
$$
\tG^2|_{\id}=-\e^{i}\e^{j}{1\/(2\n )!}\sum_{k=0}^{\n }
{2\n \choose2k}(g^{k})_{ij}(\tF^{\n -k})_{ij}\equiv-\det(g+\tF)\punkt
\AEqn\Determinant
$$

In order to complete the proof we must show that the terms in $\tG^2$ 
containing $\g$-matrices vanish. 
To this end, we return to the identity (\GammaProduct).  Not all terms
in this identity will survive; actually  
$$
\e^{i}\e^{j}g_{[i_{1}...i_{2q}}^{j_{1}...j_{2q}}
\g_{i_{2q+1}...i_{2k}j_{2q+1}...j_{2r}]}=0
\Rightarrow \e^{i}\e^{j}g_{[i_{1}...i_{2q}}^{j_{1}...j_{2q}}
\g_{i_{2q+1}...i_{2k}]j_{2q+1}...j_{2r}}=0\punkt
\AEqn\GammaSurviving
$$
This places the constraint $2q\ge 2(k+r-\n )$ on $q$, which implies that if we
look at a term of the power $2s$ in $\tF$ ($\Rightarrow 2q\ge 2\n -4s$) and the
power $2\n -4s+2k$ in $g$ , where $0\le k\le s-1$ ($k=s$ is excluded 
since this term is
already accounted for in (\GammaProductSurviving) above), we have the 
following sum to consider:
$$ 
\sum_{l=0}^{2s-2k}\a_{l}^{s,k}(g^{\n -2s+k})_{ij}\g_{i^{(4s-4k-2l)}j^{(2l)}}
(\tF^{k+l})_{ii}(\tF^{2s-k-l})_{jj}\punkt
\AEqn\GammaVanishing
$$
Here $\g_{i^{(r)}j^{(q)}}$ has $r$ $i$-indices and $q$ $j$-indices.
{}From (\GammaProductCoefficients) we can read off the
$\a_{l}^{s,k}$ coefficient
$$ 
\a_{l}^{s,k}={(-1)^{l}\/2^{2s}(2\n -4s+2k)!\,(4s-4k-2l)!\,(2l)!\,(k+l)!\,
	(2s-k-l)!}
\punkt
\AEqn\Ccoefficient
$$
In the the sum (\GammaVanishing) we move the $i$-indices in $\g$
over to $\tF$. Using the relation  
$$
\eqalign{\e^{i}&\e^{j}(g^{\n -2s+k})_{ij}\g_{i^{(4s-4k-2l)}j^{(2l)}}
(\tF^{k+l})_{ii}(\tF^{2s-k-l})_{jj}\cr
&={2(2s-k-j)\/2j+1}
\e^{i}\e^{j}(g^{\n -2s+k})_{ij}\g_{i^{(4s-4k-2l-1)}j^{(2l+1)}}
(\tF^{k+l})_{ii}(\tF^{2s-k-l-1})_{jj}\tF_{ij}\cr}
\AEqn\SchoutenII
$$
repeatedly in both directions, we obtain
$$
\eqalign{\e^{i}\e^{j}&(g^{\n -2s+k})_{ij}\g_{i^{(4s-4k-2l)}j^{(2l)}}
(\tF^{k+l})_{ii}(\tF^{2s-k-l})_{jj}\cr
&\qquad=\b_{l}^{s,k}\e^{i}\e^{j}(g^{\n -2s+k})_{ij}\g_{j^{(4s-4k)}}
(\tF^{2s-k})_{ii}(\tF^{k})_{jj}\komma\cr}
\AEqn\SchoutenIII
$$
where
$$
\eqalign{\b_{l=0}^{s,k}&=\b_{l=2(s-k)}^{s,k}=1\komma\cr
\b_{l}^{s,k}&={(2l-1)!!\,(4s-4k-2l-1)!!\,(2s-k-j)!\,(k+j)!\/(4s-4k-1)!!\,
	k!\,(2s-k)!}
\komma\quad l\neq 0,2(s-k)\punkt\cr}
\AEqn\BetaCoefficient
$$
Inserting these expressions in (\GammaVanishing) we can factor out 
$g$, $\g$ and $\tF$, 
leaving us with the sum $\sum_{l=0}^{2(s-k)}\a_{l}^{s,k}\b_{l}^{s,k}$.
Finally, by factoring out the $l$-independent part 
we find
$$
\sum_{l=0}^{2s-2k}\a_{l}^{s,k}\b_{l}^{s,k}=
A^{s,k}\sum_{l=0}^{2s-2k}(-1)^{l}{{2s-2k}\choose l}=0\komma
\AEqn\Vanishing
$$
thereby completing the proof.

\appendix{{\fourteenmath\char'24}-variation of the \DBI\ action}Below 
we give the details concerning the calculation of the $\k$-variation
of the \DBI\ action that we left out in section {\old 4}. 
For pedagogical reasons we restrict the presentation to the type 
\II B branes, since this case displays all the relevant 
features.
At the end of the appendix, we will indicate how the notation introduced 
in section {\old 4} allows for a simultaneous analysis of the \II A and 
\II B cases.  

Let us begin by recalling some results from section {\old 4}, written
here specifically for the \II B case. When convenient, we will use the
integer $\n$ defined by $2\n=p+1$. 
The variation of the \DBI\ action with parameter $\G\k$ is
$$
\d_{\G\k}I_{\dbi}=\int_M d^{p+1}\x\,\L_{\dbi}
	\left(\genfrac{{\eightmath p}-3}{4}
\d_{\G\k}\p+\half\tr\{(g+\tF)^{-1}(\d_{\G\k}g+\d_{\G\k}\tF)\}
\right)\komma
\AEqn\BDBIvariation
$$
where $\tF\equiv e^{-{\p\/2}}\F$, and
$$\eqalign{
\d_{\G\k} g_{ij}&=4i(\bar E_{(i}\g_{j)}\,\G\k)\komma\cr
\d_{\G\k} \tF_{ij}&=4i(\bar E_{[i}\g_{j]}K\,\G\k)
		+(\bar\LL(\g_{ij}K-\tF_{ij})\,\G\k)\komma\cr
\d_{\G\k}\p&=2\bar\LL\,\G\k\punkt\cr}\AEqn\BgfpWithConstr
$$
The expression (\TheGammaMatrix) for $\G$ reduces in the \II B case to
$$
d^{p+1}\x\,\G=-{e^{\tfrac{4}(p-3)\p}\/\L_{\dbi}}\,e^{\tF}\!\!\wedge 
X I|_{\rm vol}\komma
\AEqn\BTheGammaMatrix
$$
with
$$
X=\bigoplus_{n=0}^{\n}\g_{(2n)}K^{n}\punkt\aeqn
$$

The variation (\BDBIvariation) is naturally written as a sum
$\d_{\G\k}I_{\dbi}=\D_A+\D_B+\D_C$.
The first term corresponds to the first term on the right hand side of
(\BDBIvariation) and is the one that is easiest to calculate. Indeed,
we only need to insert the expression for $\d_{\G\k}\p$ from 
(\BgfpWithConstr) in (\BDBIvariation). We can then immediately use the
form expansion (\BTheGammaMatrix) to obtain 
$$
\eqalign{
\D_A&=-\frac{2}(p-3)\int_M e^{\tfrac{4}(p-3)\p}
e^{\tF}\!\!\wedge (\bar\LL XI\k)\cr
&=-\frac{2}(p-3)\int_M e^\F\!\!\wedge\bigoplus_{n=0}^{\n}e^{\tfrac{2}(n-2)\p}
(\bar\LL\g_{(2n)}K^{n}I\k)\punkt\cr}
\AEqn\DeltaA
$$
In the last step we exploited the fact that only terms proportional to the
volume form survive the integration, to move the dilaton factor in 
$\tF$ over to the $\g$-terms. Note here that the total variation
$\d_{\G\k}I_{\dbi}$ is not allowed to contain any $p$-dependent terms, if we
are to be able to recover the supergravity constraints for the 
\RR\ fields strengths by imposing $\d_{\G\k}I_{\dbi}+\d_\k I_{\wz}=0$.
The $p$-dependence in (\DeltaA) must therefore be canceled by the remaining
terms $\D_B+\D_C$. 

Having determined $\D_A$, we turn to $\D_B$ which we define as the
$\LL$-independent part of $\d_{\G\k}I_{\dbi}$. This is the part that
will eventually be canceled by the dimension-$0$-component field strength
contribution to $\d_\k I_{\wz}$. It is most conveniently calculated by
first observing that 
$$
\eqalign{
\D_B&=2i\int_M d^{2\n}\x\,\L_{\dbi}((g+\tF)^{-1})^{ji}(\bar E_{(i}\g_{j)}
\,\G\k)+ (\bar E_{[i}\g_{j]}K\,\G\k))\cr
&=2i\int_M d^{2\n}\x\,\L_{\dbi}((g+\tF K)^{-1})^{ji}(\bar E_{i}\g_{j}
\,\G\k)\komma\cr}
\AEqn{\DeltaB}
$$
due to the property $K^2=\id$. We then write 
$(\bar E_{i}\g_{j}\,\G\k)=(\g_j\,\G)_{\a\b}E_i^\a\k^\b$ and insert the
expression (\BTheGammaMatrix) for $\G$, thus finding
$$
\eqalign{
\D_B
&=-2i\int_M e^{(\tfrac{2}\n-1)\p}((g+\tF K)^{-1})^{ji}
\sum_{n=0}^{\n}\left(\frac{($\n$-$n$)!}\tF^{\n-n}\!\wedge
(\g_j\g_{(2n)}K^{n}I)_{\a\b}\right)E_i^\a\k^\b\punkt\cr}
\AEqn\DeltaBB
$$ 
By writing out the forms in components and using the 
$\g$-matrix identity 
$$
\g_j\g_{i_1...i_{2n}}=\g_{ji_1...i_{2n}}+2n g_{j[i_1}\g_{i_2...i_{2n}]}
\komma
\aeqn
$$
the sum in (\DeltaBB) above becomes    
$$
\sum_{n=0}^{\n}\left(\g_{ji_1...i_{2n}}\tF_{i_{2n+1}i_{2n+2}}
+{{2(n-k)}\over{2k+1}}g_{ji_1}\g_{i_2...i_{2n+2}}K\right)
{(\tF^{\n-n-1})_{i_{2n+3}...i_{2\n}}K^{n}\over{(\n-n)!(2n)!2^{\n-n}}}
\punkt 
\aeqn
$$
Here we have left out the factor $d^{2\n}\x\,\e^{i_{2\n}...i_1}$
and relabeled the summation index in order to pair together terms
with equal numbers of $\g$-matrices.  

Cycling the $2\n$+1 indices on the left term using $S_{[ji_1...i_{2\n}]}=0$
we can move the free index $j$ over to one of the $\tF$'s, at the
same time picking up
precisely the combinatorical factor that allows us to extract the
matrix $(g+\tF K)_{i_1j}$ (to achieve this we again need to use
$K^2=\id$). This matrix is then contracted with its inverse
appearing in (\DeltaB) to yield $\d^i_{i_1}$, and after some straightforward
steps we arrive at the result
$$
\D_B=2i\int_M e^\F\!\!\wedge\bigoplus_{n=1}^{\n} e^{\tfrac{2}(n-1)\p}
(\bar E\wedge\g_{(2n-1)}
K^n I\k)\punkt
\AEqn\DeltaBRes
$$

The third and final contribution to $\d_{\G\k}I_{\dbi}$ is the trickiest
one to compute. It can be written as
$$
\D_C=\half\int_M d^{2\n}\x\,\L_{\dbi}((g+\tF)^{-1})^{ji}
(\{(\g_{ij}K+g_{ij})-(g+\tF)_{ij}\}\G)_{\a\b}\LL^\a\k^\b\punkt
\AEqn\DeltaC
$$
Here we have added and subtracted $g_{ij}$, since we can then
perform the trace in second term to obtain an expression of exactly the 
same form as $\D_A$, only with a different coefficient. By comparing these
coefficients we can use (\DeltaA) to immediately write down the result
$$
\D_{C_{2}} = \frac{2}(p+1)\int_M e^\F\!\!\wedge\bigoplus_{n=0}^{\n}
e^{\tfrac{2}(n-1)\p}(\bar\LL\g_{(2n)}K^{n}I\k)\punkt
\AEqn\DeltaCtwo
$$   
Note that the $p$-dependent part of $\D_{C_{2}}$ cancels the one of
$\D_A$, leaving a contribution with form-level-independent coefficient
(apart from the dilaton factor). 

It remains to evaluate the contribution $\D_{C_1}$ from the term
proportional to $(\g_{ij} K+g_{ij})$ in (\DeltaC). This is done by using 
essentially the same techniques as we did when determining 
$\D_{B}$. For $\D_{C_1}$, however, the manipulations are somewhat more
intricate. In order not to lose the essential ideas behind a blur of
coefficients and indices, we will be rather schematic, leaving to the
interested reader to check explicitly that the signs and coefficients
match as promised.

It is convenient to extract the matrix 
$T_{ij}\equiv (\g_{ij}K+g_{ij})\,d^{2\n}\x\,\L_{\dbi}\G$
from (\DeltaC) (here we suppress the spinor indices).
Our strategy is as for $\D_B$ to try to rearrange the indices to obtain
a factor $(g+\tF)_{i.}$ that will cancel the inverse matrix in (\DeltaC).
In analogy with the treatment of $\D_B$ above we thus insert the expression 
(\BTheGammaMatrix), write the forms in components and expand the products of 
$\g$-matrices. For this case the relevant identity contains three terms on 
the right hand side:
$$
\g_{ij}\g_{i_1...i_{2n}}=\g_{iji_1...i_{2n}}-4ng_{[i|i_1|}
\g_{j]i_2...i_{2n}}-2n(2n-1)g_{[i|i_1|}g_{j]i_2}\g_{i_3...i_{2n}}
\AEqn{\gijIdentity}
$$
(recall that there is a factor $\e^{i_{2\n}...i_1}$ in $T_{ij}$ enforcing 
antisymmetrization in the $2\n$ indices $i_1...i_{2\n}$). 
After the proper relabeling of the summation index $n$ for the first and 
third term, $T_{ij}$ can be written
as sum of terms homogeneous in powers of $\g$-matrices. Schematically we
thus have
$$
T_{ij} \sim \sum_{n=0}^{\n}\left(
\{\g_{ij}\tF^{\n-n+1}+g_i g_j\g\tF^{\n-n-1}+g_{ij}\tF^{\n-n}\}K^n
+g_{[i}\g_{j]}\tF^{\n-n} K^{n+1}\right)\punkt
\AEqn\TijApprox
$$   
Here and in the sequel we leave out the exact coefficients of the respective 
terms as well as the contracted indices $i_1...i_{2\n}$. 
Furthermore, all $\g$'s have
$2n$ indices and we have used the property $K^2=\id$. 

The ``boundary terms'' of the sum will need some special consideration, but 
since this presents no additional complications we will not give the details
here. 
As an amusing example of how analysing boundary terms can reveal useful
information, it is worth mentioning, however, that if 
one considers $\D_C$ as a whole using instead of (\MatrixVariation) the 
identity
$$
\d\det M = {1\over{(n-1)!}}\e^{i_1...i_n}\e^{j_1...j_n}
M_{i_1j_1}...M_{i_{n-1}j_{n-1}}\d M_{i_n j_n}\komma
\aeqn
$$
one finds after expanding the $\g$-matrix products that the 
$\tF$-independent terms in the form-expansion sum must vanish 
for symmetry reasons.
This means that for fixed $p$ the $\F$-independent term in 
(\DeltaKappaWZ) containing the dimension-$\half$-component of 
$R_{(p+2)}$ is canceled solely by the $\F$-independent term in 
$\D_A$ which was very straightforward to determine. Hence we
can immediately read off the components $R_{a_1...a_{p+1}\a}$ 
from (\DeltaA) without having to calculate the more difficult 
term $\D_C$. Of course, in order to prove $\k$-invariance we 
have to go through the whole calculation, but this observation 
provides a very strong hint that we are on the right track.    

After this short excursion, let us thus return to the computation 
of $\D_{C_1}$.
We use again the trick of antisymmetrizing in $2\n+1$ indices to reposition
the indices $i$ and $j$. For the first term in (\TijApprox) one obtains in
this way
$$
\g_{ij}\tF^{\n-n+1} \sim \tF_{[i}\g_{j]}\tF^{\n-n} =
\tF_{i}\g_{j}\tF^{\n-n}\komma
\AEqn\TheFpart 
$$
where in the last step we used the observation that the symmetrized piece 
vanishes by cycling the indices $ji_1...i_{2\n}$. The term (\TheFpart) is
the $\tF$-part of the matrix $(g+\tF)_{i.}$ that we wish to extract from
$T_{ij}$. It thus remains to find the corresponding $g$-term and show
that the rest vanishes. 

The required term can be isolated from the third term in (\TijApprox);
after cycling the indices $ji_1...i_{2\n}$ the latter reads
$$
g_{ij}\tF^{\n-n} \sim g_{i}\g_{j}\tF^{\n-n} -g_{[i}\g_{j]}\tF^{\n-n} 
+ g_{(i}\tF_{j)}\g\tF^{\n-n-1}\punkt
\aeqn
$$
The first term on the right hand side is exactly the one we want;
together with (\TheFpart) it gives
$$
\D_{C_1} = -\int_M e^\F\!\!\wedge\bigoplus_{n=0}^{\n}n e^{\tfrac{2}(n-1)}
(\bar\LL\g_{(2n)}K^n I\k)
\punkt\AEqn\DeltaCone
$$
Here we have anticipated that the remaining terms 
$$
\eqalign{
\tilde T_{ij} \sim \sum_{n=0}^{\n} \left(\{g_i g_j\g\tF^{\n-n-1}
\right.&+g_{[i}\g_{j]}\tF^{\n-n}+g_{(i}\tF_{j)}\g\tF^{\n-n-1}\}K^n\cr
&\left.+g_{[i}\g_{j]}\tF^{\n-n} K^{n+1}\right)\cr}
\AEqn\TijRest
$$
in $T_{ij}$ will cancel out; we will now show that this is really the case. 

We begin by noticing that the last term has a different $K$-dependence,
and must vanish by itself. Indeed, by cycling again in $ji_1...i_{2\n}$
the index $j$ ends up only on an $\tF$-factor due to the symmetry of $g$.
By contracting the resulting $g_{[i}\tF_{j]}$ with the matrix 
$((g+\tF)^{-1})^{ji}$ one finds after some straightforward algebra that the 
term is proportional to the matrix
$$
\tF(1+g^{-1}\tF)^{-1}-(1+\tF g^{-1})^{-1}\tF\komma\aeqn
$$
which is readily found to vanish identically.   

The cancellation of the remaining three terms in (\TijRest) is a bit less
obvious; by cycling the indices $ji_1...i_{2\n}$ in the second term we 
obtain a term $g_{[i}\tF_{j]}\g\tF^{\n-n-1}$ with the right coefficient
to combined with the third term to give the term $g_{i}\tF_{j}\g\tF^{\n-n-1}$
containing no explicit \hbox{(anti)}symmetrization. When added to the first
term in (\TijRest) the latter gives us the term $g_i (g+\tF)_{.j}$, which 
can be contracted with $((g+\tF)^{-1})^{ji}$ from the left. The result
is a factor $g_{i_1i_2}$ which vanishes due to the antisymmetrization in
the indices $i_1...i_{2\n}$. This concludes the proof of (\DeltaCone).

Let us then collect the results that we have obtained; combining (\DeltaA),
(\DeltaBRes), (\DeltaCtwo) and (\DeltaCone) we find the following expression
for the $\k$-variation of the \DBI\ action for a general \II B D$p$-brane:
$$
\eqalign{
\d_{\G\k}I_{\dbi} = \int_M e^\F\!\!\wedge\bigoplus_{n}\Bigl[& 
	2i\,e^{\tfrac{2}(n-1)\p}(\bar E\wedge\g_{(2n-1)}K^n I\k)\cr 
& +(2-n)e^{\tfrac{2}(n-1)\p}(\bar\LL\g_{(2n)}K^{n}I\k)\Bigr]\punkt\cr
} 
\AEqn\DeltaDBIIIB
$$

As mentioned in the beginning of this appendix, the above calculations
can be performed quite analogously for the \II A 
case, or even simultaneously for the all D$p$-branes. 
The only essential difference between the \II A and \II B cases, 
as far as $\k$-symmetry is concerned, can be traced to the fact that 
$\gel$ anticommutes whereas $K$ commutes with the $\g$-matrices.
The notation introduced in section {\old 4} has been tailored 
specifically to deal with the consequences of this difference in an
efficient manner. 

A convenient starting-point for a simultaneous treatment of the
\II A and \II B D-branes is the following expression for the 
$\k$-variations listed in (\gfpWithConstr): 
$$\eqalign{
d^{p+1}\x\delta_{\G\k}(g+\tF)_{ij}&=
		-{e^{{p-3\/4}\phi}\/\L_{\dbi}}e^{\tilde{\F}}\!\!\wedge
		(4i \bar{E}_{(i}\g_{j)}XY\k+
		4i\bar{E}_{[i}\g_{j]}X P Y\k\cr
		&+\bar{\LL}\g_{ij}X P Y \k-
		\bar{\LL}XY\k\tilde{\F}_{ij})\punkt\cr}
\AEqn\MVariationAandB
$$
Note that for type \II A we have 
$XP=-PX$. It is however only when 
the combination $XP$ is used in (\MVariationAandB) 
that the type \II A expression is formally equivalent 
(signwise) to the type \II B expression, since
$X$ is always written with $P$ to the right in (\TheGammaMatrix).
This observation is of course irrelevant for the final answer, 
but relevant if one wishes to treat type \II A and type \II B
simultaneously. The outcome of such an analysis is the expressions
(\DeltaDBInoll) and (\DeltaDBIhalv).

\vfill\eject
\frenchspacing
\refout
\end